\documentclass[twocolumn,aps,prx,reprint,superscriptaddress,longbibliography]{revtex4-2}

\usepackage{dcolumn}    
\usepackage{bm}         
\usepackage{xfrac}      
\usepackage{physics}    
\usepackage{amsmath}
\usepackage{amssymb}
\usepackage{graphicx}
\usepackage{color}
\usepackage{hyperref}
\usepackage{xcolor}
\usepackage{lipsum}
\usepackage{tabularx}
\usepackage{booktabs}
\usepackage{multirow}
\usepackage{enumitem}
\usepackage{titlesec}
\usepackage{soul}

\hypersetup{
    colorlinks,
    linkcolor={blue!80!black},
    citecolor={blue!80!black},
    urlcolor={blue!80!black}
}

\usepackage[caption=false]{subfig}

\newcommand{\icol}[1]{
  \left(\begin{smallmatrix}#1\end{smallmatrix}\right)%
}

\begin{document}

\title{Sensing and Control of Single Trapped Electrons Above 1 Kelvin}

\author{K.~E.~Castoria}
\altaffiliation{These two authors contributed equally to this work.}
\affiliation{EeroQ Corporation, Chicago, Illinois, 60651, USA}

\author{N.~R.~Beysengulov}
\altaffiliation{These two authors contributed equally to this work.}
\affiliation{EeroQ Corporation, Chicago, Illinois, 60651, USA}


\author{G.~Koolstra}
\affiliation{EeroQ Corporation, Chicago, Illinois, 60651, USA}

\author{H.~Byeon}
\affiliation{EeroQ Corporation, Chicago, Illinois, 60651, USA}

\author{E.~O.~Glen}
\affiliation{EeroQ Corporation, Chicago, Illinois, 60651, USA}


\author{M.~Sammon}
\affiliation{EeroQ Corporation, Chicago, Illinois, 60651, USA}

\author{S.~A.~Lyon}
\affiliation{EeroQ Corporation, Chicago, Illinois, 60651, USA}

\author{J.~Pollanen}
\affiliation{EeroQ Corporation, Chicago, Illinois, 60651, USA}


\author{D.~G.~Rees}
\affiliation{EeroQ Corporation, Chicago, Illinois, 60651, USA}

\date{\today}

\begin{abstract}

Electrons trapped on the surface of cryogenic substrates (liquid helium, solid neon or hydrogen) are an emerging platform for quantum information processing made attractive by the inherent purity of the electron environment, the scalability of trapping devices and the predicted long lifetime of electron spin states. Here we demonstrate the spatial control and detection of single electrons above the surface of liquid helium at temperatures above 1 K. A superconducting coplanar waveguide resonator is used to read out the charge state of an electron trap defined by gate electrodes beneath the helium surface. Dispersive frequency shifts are observed as the trap is loaded with electrons, from several tens down to single electrons. These frequency shifts are in good agreement with our theoretical model that treats each electron as a classical oscillator coupled to the cavity field. This sensitive charge readout scheme can aid efforts to develop large-scale quantum processors that require the high cooling powers available in cryostats operating above 1 K. 

\end{abstract}

\maketitle

\section{Introduction}
\label{sec:intro}

The scaling of solid-state quantum information processing technologies to large qubit numbers presents a significant technological challenge as the heat loads generated by qubit operation and from external control lines increase significantly with the qubit number~\cite{hornibrook2015cryogenic, almudever2017engineering, krinner2019engineering}. This presents a particular problem for superconducting qubits~\cite{kjaergaard2020superconducting} and other systems that leverage superconducting circuit quantum electrodynamics (cQED)~\cite{clerk2020hybrid}, which typically operate in dilution refrigerators that offer only limited cooling power ($\sim$1~mW at 100~mK)~\cite{anferov2024superconducting}. Other technologies, such as electron spin qubits in silicon quantum dots, have shown more promise for operation above 1~K~\cite{yang2020operation, huang2024high}. This allows operation in pumped $^4$He cryostats in which cooling powers can exceed 100~mW~\cite{piegsa2017high, wang2018high}. For such devices, integrated single electron transistors or dispersive radio-frequency (RF) gate reflectometry techniques are commonly used for charge readout~\cite{crippa2019gate, liu2021radio, ares2016sensitive}.

Electrons trapped on cryogenic substrates have been proposed as good candidates for charge or spin qubits~\cite{schuster2010proposal,platzman1999quantum,lyon2006spin}. Several recent experiments have demonstrated that single electrons can be isolated on the surface of thin (0.01 - 1~$\mu$m) layers of liquid helium or solid neon using gate electrodes beneath the cryogen surface~\cite{papageorgiou2005counting,koolstra2019coupling, zhou2022single}. The set of gate electrodes also comprise superconducting resonators engineered to couple to the electron orbital state, the frequency of which ($\sim$5~GHz) can be tuned by the electrode bias voltages. The anharmonicity of the trapping potential ensures that the electron acts as a two-level system interacting with the microwave cavity, which can be described by the Jaynes-Cummings model familiar from cQED~\cite{wallraff2004strong}. As is standard in cQED experiments, where coherent control requires the suppression of thermal excitations comparable with the operating frequencies ($2–10$~GHz), these electron trapping experiments were conducted close to 10 mK. Large shifts in the cavity frequency are observed as the electron motion is tuned into resonance by varying the confining potential, with an electron-photon coupling strength $g/2\pi \approx 5$~MHz exceeding the bare cavity linewidth by a factor of 10. However, to date, the electron linewidth $\Gamma_e/2\pi \approx 80$~MHz has been limited by interactions between the electron and the helium substrate~\cite{koolstra2019coupling}. Replacing the liquid helium with a thin ($\sim10$~nm) layer of solid neon was found to result in a reduced electron linewidth ($\Gamma_e/2\pi \approx 0.36$~MHz) allowing demonstration of strong coupling and quantum gate operation~\cite{zhou2022single, zhou2024electron}. However, the long charge qubit relaxation time ($T_1 = 48$~$\mu$s) and phase coherence time ($T_2 = 93$~$\mu$s) are achieved at the cost of severely reduced trapping reproducibility due to the localization of electrons by random surface roughness~\cite{troyanovskii1979electron, kajita1982two}.  

In contrast to the orbital state, the spin state of an electron trapped on helium is expected to show long coherence times and remain robust even at temperatures exceeding 1 K~\cite{lyon2006spin, dykman2023spin}. Because these electrons are identical, move with record high mobility in a defect-free environment~\cite{grimes1976observation}, and can be manipulated in micron-scale traps with dc gate electrodes~\cite{rees2011point}, a system of such qubits should be readily scalable to large numbers. Furthermore, spin-orbit coupling induced by local magnetic field gradients should allow readout of the electron spin state via CPW resonators coupled to the electron motion~\cite{schuster2010proposal, kawakami2014electrical}. As a first step towards a spin-based processor, the trapping and efficient clocked transport of charge `packets' containing small numbers of electrons has been demonstrated in CMOS-fabricated chips covered with a helium dielectric layer~\cite{bradbury2011efficient, takita2012spatial}. However, electron detection in these devices has so far been performed via image current measurements, the sensitivity of which has so far proved insufficient to allow single electron detection. There is therefore significant interest in developing technologies for the control and sensing of single surface-state electrons that can be integrated into different device architectures and operated at elevated temperatures.
 
In this work, we demonstrate the detection of single electrons on helium at temperatures above 1~K using a CPW resonator with integrated electron trap. At these temperatures, the thermal energy exceeds the electron motional frequency by several times, and the $^4$He vapor pressure is non-negligible creating a substantially different electron environment. 
Despite these conditions we resolve reproducible resonator frequency shifts as our electron trap is loaded with single electrons. Compared with previous electron-on-helium cQED experiments at 10~mK, our single electron-resonator coupling is twice as strong and the resonator is completely decoupled from the larger reservoir of electrons. We compare measured frequency shifts with a classical model of the many electron-resonator coupling in which the motion of the trapped electrons modifies the polarization of the trap volume, thereby increasing the resonator capacitance and reducing its resonant frequency. This model, in combination with finite-element modeling (FEM) of the trapping potential, reproduces the experimental frequency shifts to good accuracy. Our results lay the ground work for exploring single electron-on-helium spin properties at elevated temperatures and can aid efforts to develop large scale quantum processors with electrons trapped at the surface of noble gas substrates~\cite{platzman1999quantum, lyon2006spin, jennings2024quantum}, as well as electrons in radio-frequency traps~\cite{daniilidis2013quantum} or in semiconductor quantum dots~\cite{burkard2020superconductor}. 

The paper is arranged as follows: in Section~\ref{sec:readout} we describe the device used in the experiment and give details of the measurement set-up and charge readout scheme. In Section~\ref{sec:experiments} we present experimental results demonstrating the controlled loading and unloading of the electron trap, as well as deterministic single electron control, and compare the observed frequency shifts with those predicted by the model. In Section~\ref{sec:discussion} we draw our conclusions.  

\section{Electron Trap and Readout Scheme}
\label{sec:readout}

\begin{figure}
\includegraphics[scale=0.22]{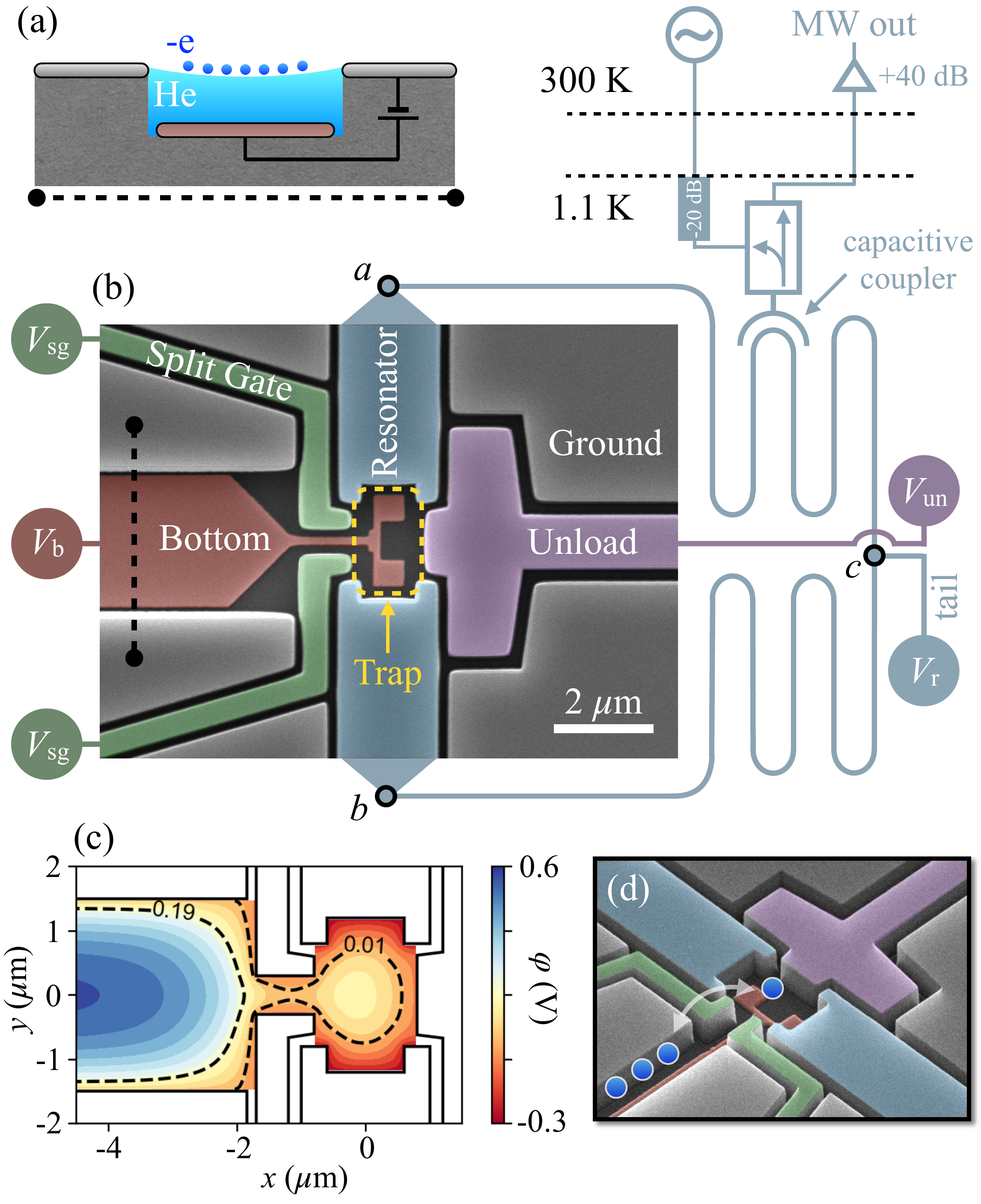}
\caption{ (a) Schematic cross-section of a microchannel filled with superfluid helium. The helium surface is charged with a system of electrons held on the surface by a potential difference applied between the upper and lower electrodes. The dashed line corresponds to that shown in (b), which indicates the path of the cross-section. (b) Schematic circuit diagram and false-color SEM image of the electron trap. For the differential resonance mode the ac voltage components at points $a$ and $b$ oscillate out of phase, with a voltage node established at point $c$. (c) Results of FEM calculation of the electrostatic potential $\varphi(x,y)$ in the plane of the helium surface. The dashed lines show isopotentials for 0.01 V and 0.19 V, as labeled. The calculation is performed for typical trapping voltages: $V_{\mathrm{b}}= 0.5$~V, $V_{\mathrm{sg}}= -0.2$~V, $V_{\mathrm{r}}= -0.3$~V and $V_{\mathrm{un}}= -0.5$~V. (d) Schematic depiction of the loading and unloading of the electron trap from the adjacent microchannel.}
\label{fig:device}
\end{figure}

All measurements were conducted in a continuous-flow $^4$He cryostat (ICEOxford DRYICE 1.0K) at a stable base temperature of 1.10~K. The device used in the experiments is composed of a long microchannel, 3~$\mu$m wide and 700~nm deep, which terminates at the electron trap. The microchannel and trap are filled with superfluid helium, the surface of which is charged with electrons, as shown schematically in Fig.~\ref{fig:device}(a). The microchannel therefore holds an extended reservoir of electrons from which a small number can be loaded into the trap as depicted in Fig.~\ref{fig:device}(d). The separation of the electron reservoir from the readout resonator provides a clear advantage over previous devices in which both the electron trap and reservoir regions were coupled to the resonator~\cite{yang2016coupling, koolstra2019coupling}. Further details of the device are given in Appendix~\ref{sec:device}. 

The electron trap is formed by the set of electrodes shown in Fig.~\ref{fig:device}(b) which are arranged to create a controllable potential well. In the plane of the helium surface the Split Gate electrodes (bias voltage $V_{\mathrm{sg}}$) create a potential barrier between the trap and the electron reservoir, while the Unload Gate (bias voltage $V_{\mathrm{un}}$) and the upper and lower arms of the resonator electrode (bias voltage $V_{\mathrm{r}}$) provide electrostatic confinement. In the lower plane, at the bottom of the microchannel, the Bottom electrode (bias voltage $V_{\mathrm{b}}$) extends into the base of the trap region where it provides an attractive potential beneath the electrons. FEM is performed to calculate the electrostatic potential $\varphi(x,y)$ in the plane of the helium surface across the trap region~\cite{zhk,freefem} as well as the density of electrons $n_s$ in the microchannel. As shown in Fig.~\ref{fig:device}(c), with suitable dc bias applied to each of these electrodes an electrostatic potential minimum for electrons is formed at the trap center, separated from the adjacent electron reservoir by a potential barrier.

As well as providing dc confinement, the resonator electrode forms part of the microwave circuit used to detect electrons in the trap. The CPW resonator (cavity) is designed to support a differential mode at bare resonance frequency $f_r = 6.04383$~GHz (see details in Appendix~\ref{sec:resonator}). This particular resonator mode is described by out-of phase ac voltage oscillations at the open ends (points $a$ and $b$ in Fig.~\ref{fig:device}(b)), where the electron trap is located. Voltage oscillations across the resonator electrodes produce an oscillating electric field that drives electron motion. The resulting collective electron motion within the trap alters the resonance properties of the cavity, which we detect by measuring shifts in the resonance frequency $\Delta f$ from the bare resonance frequency. In this setup, a probe microwave signal is introduced through a small capacitive coupling port into the cavity (see Fig.~\ref{fig:device}(b)), and the reflected signal is measured.

The influence of electrons in the trap on the resonator can be understood as follows: The collective oscillations of charges within the trap can be represented as polarized matter in the cavity, with electrical susceptibility $\chi_e(\omega)$, leading to changes in the dielectric function of the medium between the two resonator arms. This results in a shift in the resonance frequency of the cavity given by $\Delta f = -f_r \Re{\chi_e(2\pi f_r)/2}$. The electric susceptibility $\chi_e(\omega)$ characterizes the system's response to a perturbing microwave field and can be described as a sum of Lorentzian responses of discrete \textit{vibrational} modes arising in the system of electrons. We find this quantity from the coupled equations of motion for electrons in the trapping potential $\varphi(x,y)$ under the external driving field (see details in Appendix~\ref{sec:ehe-res_coupling}) and it takes the form:
\begin{equation}
    \label{eq:susceptibility_main} 
    \chi_e(\omega) = \sum_{n} \frac{4g_n^2}{\omega_n^2 - \omega^2 + 2i\omega\Gamma_n},    
\end{equation}
where $g_n$ represents the coupling energy between the microwave field and the $n^{\mathrm{th}}$ vibrational mode of the electron system, $\omega_n$ is the eigenfrequency of that mode, and $\Gamma_n$ is the phenomenological damping constant. In order to calculate $\Delta f$ for a given trapping potential and arbitrary number of electrons in the trap $N_e$ we first find the equilibrium configuration of electrons by energy minimization procedures. Subsequently, the frequencies and eigenvectors of the vibrational modes are obtained for a given electron configuration by numerically diagonalizing the total Hamiltonian of the electron system~\cite{koolstra2024high,quantumelectron}. The coupling energy $g_n$ is calculated based on the distribution of the electric field created by the resonator electrodes and the eigenvectors of the vibrational modes, completing the calculations of the resonator frequency shift. More details on the many electron-resonator coupling appear in Appendix~\ref{sec:ehe-res_coupling}.

\section{Experimental Results}
\label{sec:experiments}

\subsection{Trap Loading and Unloading}
\label{sec:loading}

\begin{figure}
\includegraphics[scale=0.24]{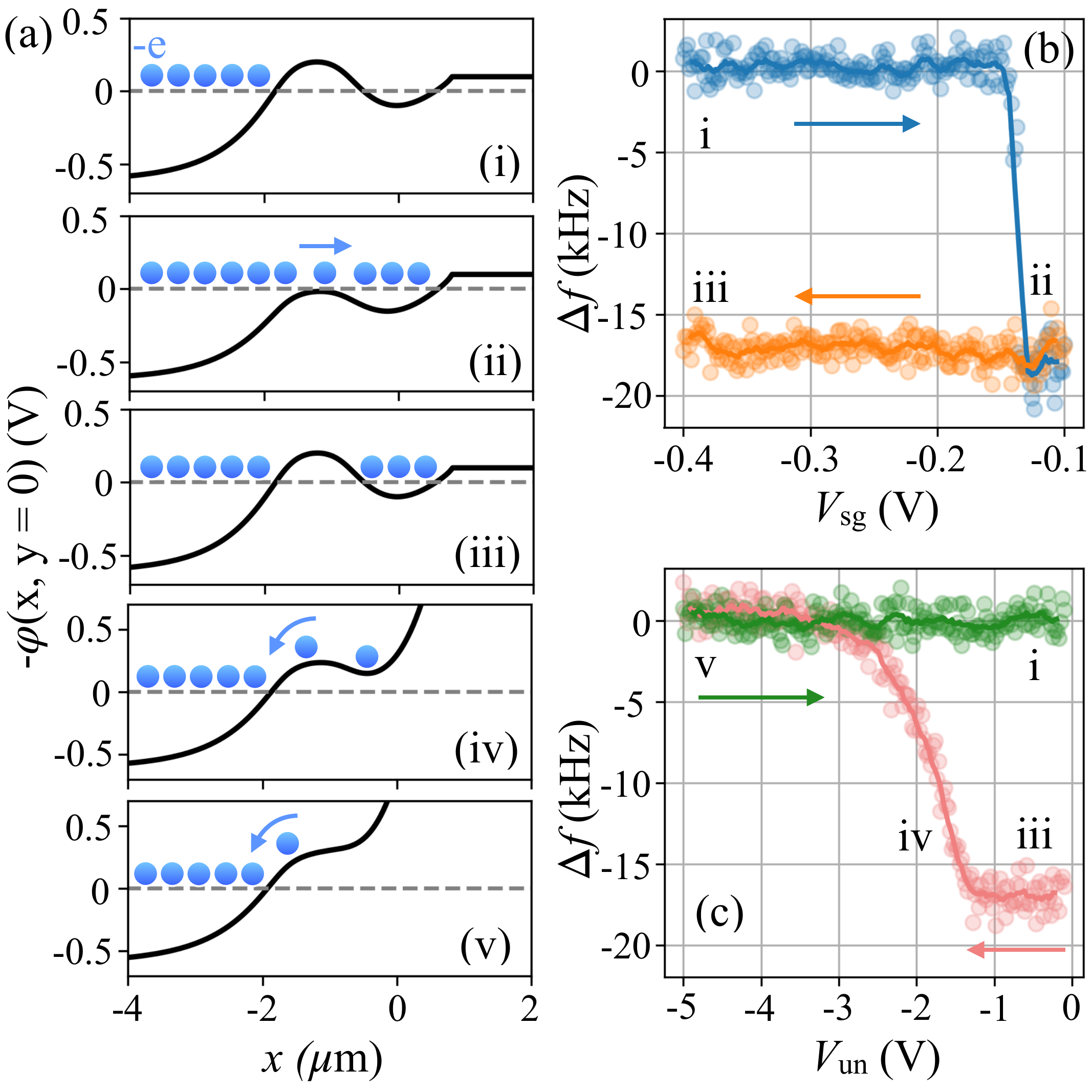}
\caption{\label{fig:loading} Loading and unloading the electron trap at $T=1.1$~K. For all plots $V_\mathrm{b} = 1.0$~V and $V_\mathrm{r} = -0.3$~V. (a) FEM analysis showing the evolution of the potential profile $-\varphi(x,0)$ during the sequence of voltage sweeps used to load and unload the electron trap. To load the trap $V_\mathrm{sg}$ is swept from -0.4~V (i) to -0.1~V (ii) and back to -0.4~V (iii) with $V_\mathrm{un}=-0.1$~V. To unload the trap $V_\mathrm{un}$ is swept to more negative voltage, with $V_\mathrm{sg}$ held at -0.4~V. For $V_\mathrm{un}=-2.0$~V (iv) the trap potential is raised and some number of the trapped electrons escape to the adjacent reservoir, and for $V_\mathrm{un}=-5.0$~V (v) the trap is emptied completely. Sweeping $V_\mathrm{un}$ back to -0.1~V returns the system to the initial condition (i). The measured $\Delta f$ during loading and unloading is shown in (b) and (c), respectively. A 10-point rolling average is added as a solid line for clarity. $\Delta f$ decreases when the trap is populated with electrons, in agreement with the electron-resonator coupling model.}
\end{figure}

Before investigating the electron-resonator coupling for a single electron we first demonstrate the controlled loading and unloading of a large number (several tens) of electrons into and out of the trap. As described in Section II, the electron trap is separated from the reservoir of electrons stored in the adjacent microchannel by a potential barrier formed between the Split Gate electrode. We perform a sequence of gate sweeps that systematically reduce the barrier height allowing electrons to spill into the trap, before raising the barrier again to corral a small number of charges within the trap (trap loading). The trap potential is then raised by sweeping the Unload gate negative in order to push electrons back over the barrier until no electrons remain (trap unloading). 

These loading and unloading procedures are depicted schematically in Fig.~\ref{fig:loading}(a) with the aid of potential profiles generated by FEM. The corresponding experimental data is shown in Fig.~\ref{fig:loading}(b) and~\ref{fig:loading}(c) for trap loading and unloading, respectively. For the initial voltage configuration ($V_{\mathrm{b}} = 1.0$~V, $V_{\mathrm{sg}} = -0.4$~V, $V_\mathrm{r} = -0.3$~V and $V_{\mathrm{un}} = -0.1$~V) a local potential minimum is formed in the trap region, separated from the microchannel region by a potential barrier (i). Because the barrier height exceeds the value of the electron potential $V_\mathrm{e}$, which is close to the ground potential 0~V, electrons cannot enter the trap. To allow electrons into the trap, $V_\mathrm{sg}$ is swept to the more positive value of -0.1~V. The height of the barrier is thereby reduced until it falls below $V_\mathrm{e}$, allowing electrons to spill into the trap (ii). Because the trap area is negligible compared to that of the adjacent reservoir microchannel network, the electrochemical potential of the electron system remains unchanged and the electrons self-arrange within the trap geometry in order to maintain electrostatic equilibrium with the extended electron system in the microchannel. In agreement with the electron-resonator coupling model presented in Section~\ref{sec:readout}, $f$ decreases as the trap is loaded with electrons. Once electrons enter the trap, $V_\mathrm{sg}$ is swept back to its initial value of -0.4~V. The potential barrier is restored, separating the trap from the microchannel, with some number of electrons remaining in the trap (iii). This completes the loading process. 

Using the potential profile generated by FEM, numerical energy minimization procedures are employed to find the equilibrium electron positions within the trap~\cite{quantumelectron}. For the voltage sweeps shown in Fig.~\ref{fig:loading}, this analysis gives the total number of electrons loaded into the trap as $N_\mathrm{e}\approx30$. This number is reasonable given the typical surface electron density $n_s \sim 10^9$~cm$^{-2}$ and the diameter of the trap potential well of approximately 1~$\mu$m. For these relatively large bias voltages the FEM analysis gives the trap curvature $\omega_e/2 \pi \approx 50$~GHz, much higher than bare resonator frequency. For these values of $N_\mathrm{e}$ and $\omega_\mathrm{e}$, our numerically calculated value of $\Delta f = -14$~kHz is in good agreement with the observed frequency shift of approximately 17~kHz.

Unloading is then performed by sweeping $V_\mathrm{un}$ negative (Fig.~\ref{fig:loading}(c)). As the electrochemical potential of the electrons in the trap rises to the level of the potential barrier, electrons escape to the microchannel and $f$ increases (iv). For $V_\mathrm{un}=-5.0$~V no potential minimum exists in the trap and all the electrons are expelled, and $\Delta f$ returns to 0~kHz (v). Sweeping $V_\mathrm{un}$ back to its initial value of $-0.1$~V returns the system to the initial configuration with the trap empty (i). The loading and unloading procedures can then be repeated. We note that $\Delta f$ decreases abruptly on electron loading because the relatively deep potential well is suddenly connected to the electron reservoir and quickly fills with electrons. In contrast, $\Delta f$ changes smoothly during unloading because the depth of the trap and therefore the number of electrons within it decreases smoothly with $V_\mathrm{un}$.

\subsection{Deterministic Electron Control}

\begin{figure*}
\includegraphics[scale=0.25]{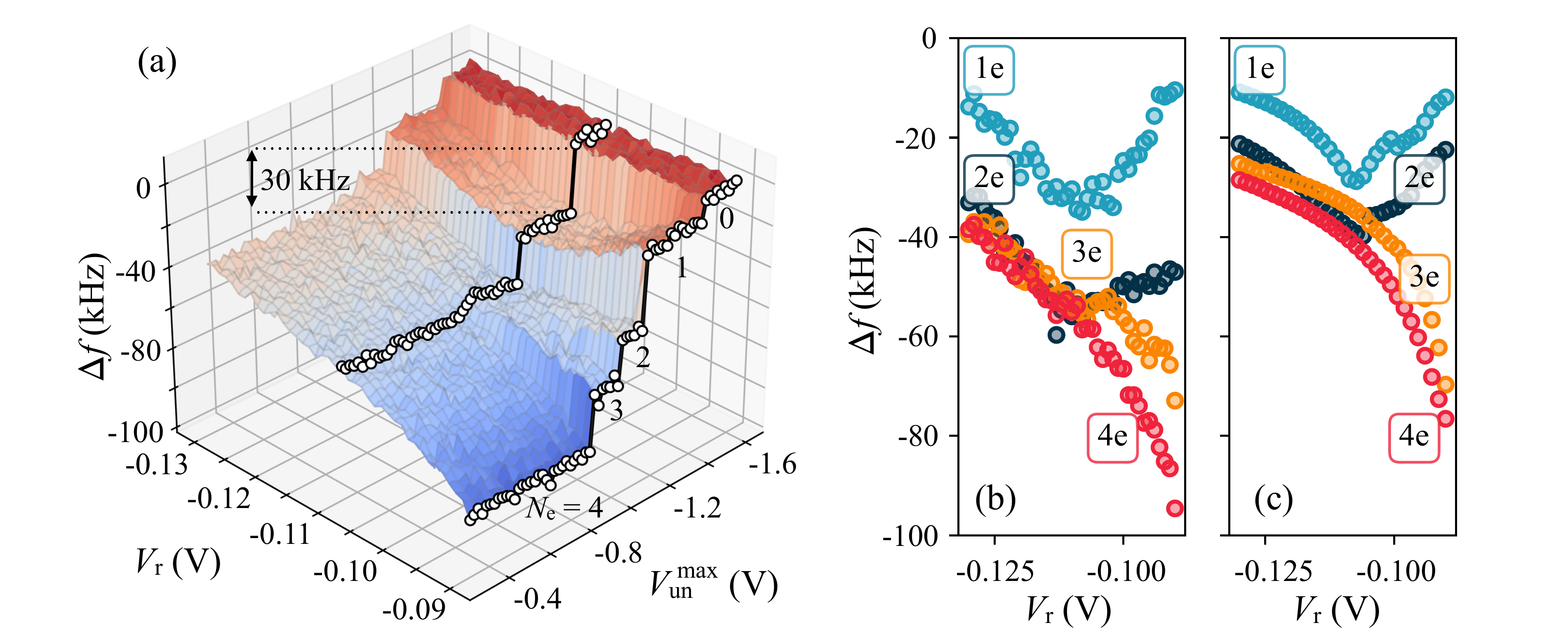}
\caption{\label{fig:multielectron} (a) Surface plot of $\Delta f$ with varying $V_{\mathrm{r}}$ and $V_\mathrm{un}^\mathrm{max}$. Two data sets corresponding to $V_\mathrm{un}^\mathrm{max}$ sweeps at constant $V_{\mathrm{r}}$ are highlighted (white circles) for $V_\mathrm{r}=-0.09$~V and $V_\mathrm{r}=-0.11$~V, in order to show clearly plateaus in $\Delta f$ corresponding to discrete $N_\mathrm{e}$ values and the maximal frequency shift for single electron trapping. (b) Slices of $\Delta f$ along $V_\mathrm{r}$ for the $N_\mathrm{e}=1, 2,3,4$ plateaus in (a). (c) $\Delta f$ calculated using the expression for the electric susceptibility $\chi_e(\omega)$, for the same conditions as in (b).}
\end{figure*}

Having demonstrated the controlled loading of the trap with several tens of electrons, we now reduce the trapping bias voltages in order to isolate a single electron. We also demonstrate that careful tuning of the confinement potential is required to bring the electron motion close to resonance with the cavity field in order to maximize the resulting $\Delta f$.

In Fig.~\ref{fig:multielectron}(a) we show $\Delta f$ as electrons are unloaded from the trap for $V_\mathrm{b}=0.3$~V. These measurements are performed in a similar manner to the loading and unloading curves shown in Fig. 3 with one important modification: after setting the Unload Gate to each increasingly negative voltage (which we denote $V_\mathrm{un}^\mathrm{max}$) the value is returned to the initial voltage of $V_\mathrm{un}=-0.4$~V. $\Delta f$ is then measured as $V_{\mathrm{r}}$ is swept between -0.09~V and -0.13~V in order to vary the curvature of the trapping potential. Returning the Unload Gate bias to the same value while $V_{\mathrm{r}}$ is swept ensures that the resonator frequency is always measured under the same bias conditions, eliminating any dependence of the electron-resonator coupling on the position of the electron system in the $x$,$y$ plane.

As in Fig.~\ref{fig:loading}(c), sweeping $V_\mathrm{un}^\mathrm{max}$ negative reduces the number of electrons in the trap. However, for this reduced bias configuration and smaller voltage step size, discrete and irreversible shifts in $\Delta f$ are observed, which we attribute to changes in the number of trapped electrons. As a result, four plateaus in $\Delta f$ are resolved, with the last plateau corresponding to a single isolated electron. However, the shape of each plateau depends strongly on the confinement curvature. For $N_\mathrm{e}=1$, the absolute value of $\Delta f$ reaches a maximum of $\sim-30$~kHz for $V_\mathrm{r}=-0.11$~V. As will be shown below, the maximum in $|\Delta f|$ corresponds to the case in which the trap curvature $\omega_e/2 \pi$ is closest to the resonance frequency of the cavity.

The experimentally measured $\Delta f$ exhibits two qualitatively distinct behaviors at the extreme values of $V_\mathrm{r}$ as the number of electrons in the trap is varied. For $V_\mathrm{r} = -0.09$~V large discrete shifts in $\Delta f$ are observed. In contrast, for $V_\mathrm{r} = -0.13$~V discrete shifts are not clearly observed for $N_e > 2$. Our eigenmode analysis of the electron system indicates that the primary modes contributing to the electric susceptibility for $N_e = 2, 3, 4$ originate from the in-phase motion of all electrons (in-phase-mode). For $N_e = 3 (4)$ we find that the frequency of the in-phase-mode approaches the resonator frequency $\omega_r$ at $V_\mathrm{r} \approx -0.09 (-0.08)$~V, resulting in the significant frequency shifts observed in the experiments as we vary the number of electrons in the dot. For $V_\mathrm{r} = -0.13$ V, the mode frequencies are detuned (13.6~GHz for $N_e=3$ and 14.4~GHz for $N_e=4$) significantly thus reducing the system response. This results in only small changes in $\chi_e$ as $N_e$ is varied. Our calculations of $\Delta f$ based on Eq.~\ref{eq:susceptibility_main} reproduce the observed data with reasonable accuracy, as shown in Fig.~\ref{fig:multielectron}(b) and (c).

Interestingly, the dependence of $\Delta f$ on $V_\mathrm{r}$ for two electrons exhibits a similar response as the single electron case. We measure values of $\Delta f$ that are twice as large as those in the single-electron case for $V_\mathrm{r} < -0.11$~V. This aligns with the predictions of our model which, for a harmonic approximation of the trap potential, indicates that the electron-resonator coupling energy for the in-phase motion of two electrons scales as $g_{2+} = \sqrt{2} g_\mathrm{e}$, with a frequency equal to that for the single electron case (see details in Appendix~\ref{sec:a-twoelectron}). This approximation is more accurate for $V_\mathrm{r} < -0.11$~V, where our FEM calculations predict a trap potential with a single minimum. For $V_\mathrm{r} > -0.11$~V, however, the FEM results indicate that the single-well potential transforms into a double-well structure, introducing significant nonlinear terms into the potential. As a result, an exact factor of 2 is no longer expected, consistent with our experimental observations. It is important to note that the quantum Dicke and Tavis-Cummings models predict a characteristic $\sqrt{N}$ scaling in coupling energy only under the assumption of homogeneous coupling~\cite{frisk2019ultrastrong}. In contrast, our experiments effectively realize a many-atom cavity QED system with inhomogeneous atom-cavity coupling.

\begin{figure}
\includegraphics[scale=0.24]{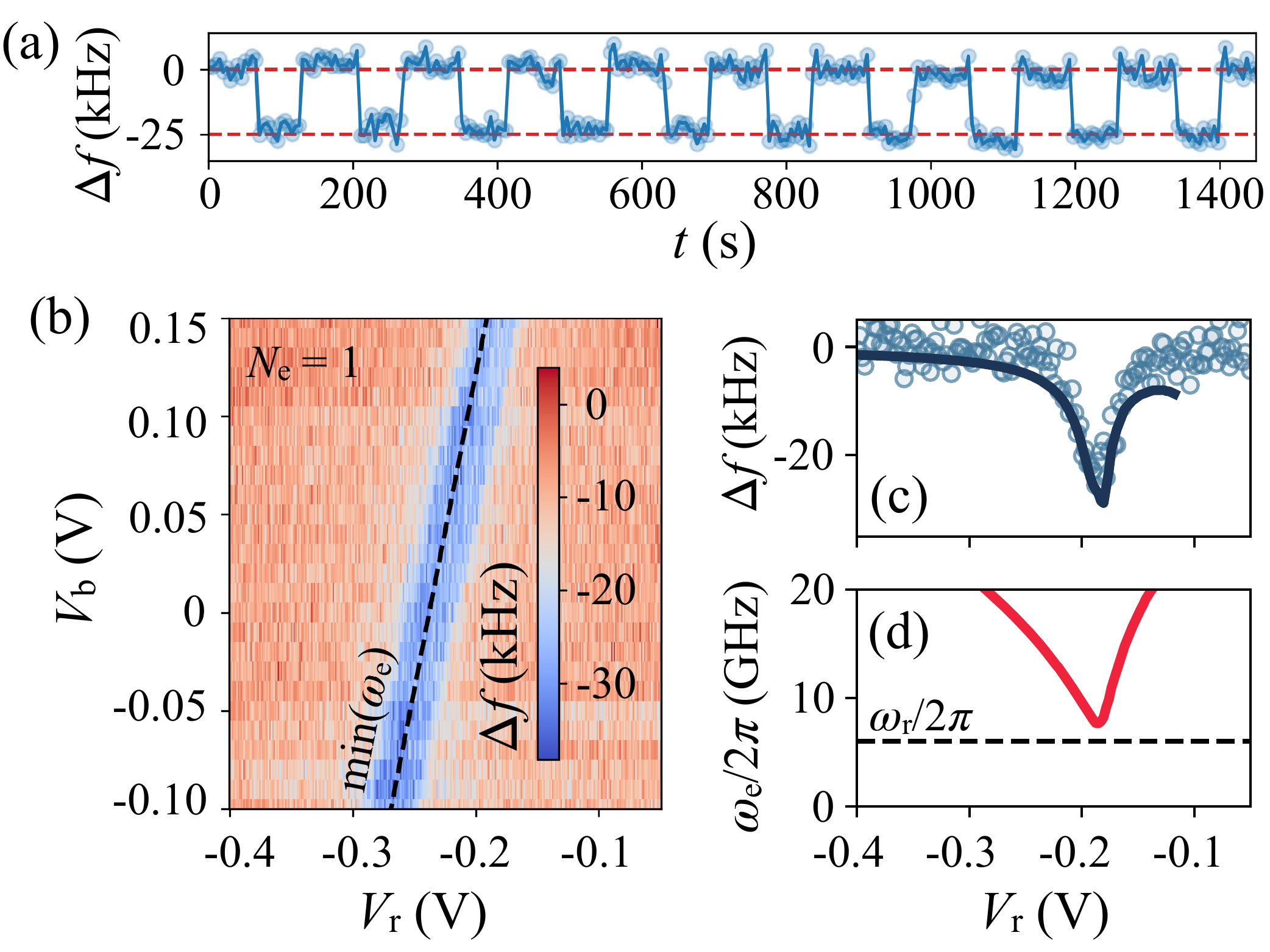}
\caption{\label{fig:singleelectron} (a) $\Delta f$ recorded over time $t$ during repeated loading and unloading of a single electron. All data points are recorded for the voltage configuration $V_\mathrm{b}=0.3$~V, $V_\mathrm{r}=-0.1$~V, $V_\mathrm{un}=-0.4$~V and $V_\mathrm{sg}=-0.5$~V. Electron loading and unloading is performed by applying a series of voltage step sequences on a timescale shorter than the measurement period (and therefore between data points). For electron loading $V_\mathrm{sg} = $ -0.4~V $\rightarrow$ -0.05~V $\rightarrow$ -0.4~V with $V_\mathrm{r} = V_\mathrm{sg}=-0.4$~V and for unloading $V_\mathrm{un} = $ -0.4~V $\rightarrow$ -3.5~V $\rightarrow$ -0.1~V with $V_\mathrm{r} = V_\mathrm{sg}=-0.4$~V. (b) $\Delta f$ against $V_\mathrm{r}$ and $V_\mathrm{b}$ for $N_\mathrm{e}=1$. Here $V_\mathrm{un}=-0.4$~V and $V_\mathrm{sg}=-0.5$~V. The voltage configurations for which $\omega_\mathrm{e}$ reaches its minimum are given by the dashed line, with a small +50~mV offset added to $V_\mathrm{r}$ to aid comparison with the experimental result. (c) $\Delta f$ against $V_\mathrm{r}$ for $V_\mathrm{tr}=0.15$~V. The solid lines are generated from Eq.~\ref{eq:susceptibility_main} using potential profiles generated by FEM, as described in the text. (d) The trap curvature $\omega_\mathrm{e} / 2\pi$ for the same conditions as in (c).}
\end{figure}

In Fig.~\ref{fig:singleelectron}(b) we demonstrate the repeated loading and unloading of the trap with a single electron. During the measurement, the bias values are chosen to give the maximal shift in $|\Delta f|$, resulting in repeated $\sim$25 kHz shifts with each loading event (the bias configurations employed are described in the caption). We observe no errors in this single electron control cycle. This measurement confirms reproducible single electron control with good detection sensitivity. This is unsurprising given the helium surface is free of defects and that the energy required to escape the potential well (which is 60~mV deep according to FEM) greatly exceeds the electron temperature ($k_{\mathrm{B}}T/e\approx0.1$~mV). We note that these single electron measurements are made without optimization of data acquisition speed. In addition, without amplification at the cryogenic stage our measurement is limited by noise from the room temperature amplifiers. Therefore, significant improvements in detection speed and sensitivity can be achieved in future experiments, which may be aided by more advanced measurement techniques such as homodyne sideband detection under gate voltage modulation of the electron signal~\cite{vigneau2023probing}.  

The dependence of $\Delta f$ on $V_{\mathrm{r}}$ occurs due to the variation of the confinement curvature, and according to Eq.~\eqref{eq:susceptibility_main} the maximum shift in $\Delta f$ occurs when the single-electron frequency $\omega_\mathrm{e}$ approaches closest to the resonator frequency $\omega_r$. As the trap potential continuously evolves from a single well to a double well structure, we expect the trap curvature $\omega_\mathrm{e}$ to decrease, reaching a minimum at certain values of $V_\mathrm{r}$ and $V_b$, and then increase. We confirm this in Fig.~\ref{fig:singleelectron}(c) by measuring $\Delta f$ while varying $V_{\mathrm{r}}$ and $V_{\mathrm{b}}$ for $N_e=1$. Because the curvature of the confinement potential depends on both voltages the minimum in $\Delta f$ moves along a linear path in the $V_{\mathrm{r}} - V_{\mathrm{b}}$ plane. We also show the line of minimum $\omega_\mathrm{e}$, calculated using FEM, which closely follows the experimental dip. The good agreement confirms that $\omega_\mathrm{e}$ is accurately controlled by tuning the confining bias voltages. In Fig.~\ref{fig:singleelectron}(c), we present the same frequency shifts as a function of $V_\mathrm{r}$, measured at fixed $V_{\mathrm{b}}$, alongside predictions from our model based on Eq.~\ref{eq:susceptibility_main} (solid curve) with single electron coupling strength $g_\mathrm{e}/2\pi = 9.8$~MHz, showing good agreement. The calculated values of $\omega_\mathrm{e}$ for these voltage configurations are also shown in Fig.~\ref{fig:singleelectron}(d). We note that the value of $g_\mathrm{e}$ for our device is a factor of two larger than those previously reported in cQED experiments with electrons on helium~\cite{koolstra2019coupling}. The observation of a single minimum in the frequency shift suggests that $\omega_e > \omega_r$ as the potential transitions from a single-well to a double-well structure. We find that this scenario is only possible if a small uncompensated electric field exists along the $y$-axis (modeled in our study by applying a 20~mV difference between the two split-gate electrodes). Such a field could arise from a slight misalignment of the upper and lower electrode layers during the fabrication process.

\section{Discussion and Conclusions}
\label{sec:discussion}

We have demonstrated the implementation at temperatures above 1 K of a CPW resonator as a charge sensing element for single electrons trapped on the surface of liquid helium. While the device and measurement circuit are familiar from cQED experiments, we have developed a classical description of the resonator-electron coupling for this elevated temperature that accurately predicts the observed frequency shifts for both large electron ensembles and single electrons. The shift in the resonance frequency can be understood as a renormalization of the resonator capacitance by an effective \textit{single-electron capacitance}, analogous to the concept of quantum capacitance~\cite{Park_Metzger_Tosi_Goffman_Urbina_Pothier_Yeyati_2020}, defined as $\delta C_e = \chi_e(\omega) C \propto (\omega_e^2 - \omega_r^2)^{-1}$ in the dispersive regime, where  $C$ is the total capacitance of the resonator. For the largest frequency shift shown in Fig.~\ref{fig:singleelectron}(c), we estimate $\delta C_e \approx 0.45$~aF. To contextualize this value, we compare it to other measurement techniques. For instance, $\delta C_e \approx 10^{-3}$~aF was estimated for a single electron's out-of-plane motion from many electron image-charge Rydberg state detection methods~\cite{kawakami2019image}. For single electrons in semiconductor quantum dots $\delta C_e \approx 1-10$~aF has been reported~\cite{gonzalez2015probing,colless2013dispersive} using rf reflectometry. Our measured values of $\delta C_e$ can be improved further by bringing the electron motion fully into resonance with the cavity field, bringing our measurement technique to the level of the most advanced rf-SET charge sensors. Additionally, our theoretical model can be applied generally to LC resonator circuits capacitively coupled to bound electrons, which have been proposed for a variety of quantum computing architectures~\cite{kawakami2023blueprint, belianchikov2024cryogenic, jennings2024quantum, daniilidis2013quantum}.

The ability to detect single electrons and electron clusters at elevated temperatures could open investigations of finite size effects on phase transitions from disordered, liquid-like states, to Wigner molecule formations~\cite{rees2016structural,koolstra2019coupling}. For electron clusters, the characteristic plasma parameter, which quantifies the ratio of average Coulomb energy $\propto \sqrt{n_s}$ to thermal energy $\propto T$ of electrons, can be adjusted by varying the temperature or the confining potential. The simplest scenario in which Coulomb interactions influence the properties of an electron system can be explored and tested is the case of two electrons. While in our experiments the in-phase motion of two electrons in the trap is coupled to the differential mode of the resonator, in general the out-of-phase motion of two electrons can be driven by the common mode. Therefore, the accurate engineering of the resonance frequency of both the common and differential modes of the resonator within the measurement frequency band could allow investigation of correlated electron motion driven by Coulomb interactions. Operating at low temperatures ($k_{\mathrm{B}} T \ll \hbar \omega_n$) and in the strong coupling regime ($g_n \geq \Gamma_n$) would make it possible to study entanglement in a two-electron system~\cite{beysengulov2024coulomb}.
The level of electron control demonstrated in this work, together with engineering strong coupling between a many-body electron system and a cavity mode~\cite{koolstra2024high}, could provide a new platform for exploring quantum optics models of light-matter interactions. This includes models such as the quantum Rabi, Dicke~\cite{dicke1954coherence} and Hopfield~\cite{hopfield1958theory} models, along with newer frameworks that extend beyond these traditional models~\cite{frisk2019ultrastrong,rokaj2022free}.

In our experiments, we did not observe statistically significant changes in the resonator linewidth that would allow direct measurement of the damping rates of the electron motion. However, we found that using $\Gamma_n / 2\pi \approx 1-2$~GHz provides more accurate agreement between the experiment and our model. Studies including electron mobility measurements~\cite{Mehrotra_Guo_Ruan_Mast_Dahm_1984} and Rydberg state spectroscopy~\cite{Kawakami_Elarabi_Konstantinov_2021} have shown helium gas atom scattering to be the dominant relaxation mechanism for 2D electrons on helium for $T \gtrsim 0.7$~K. A quantum-mechanical description of relaxation rates in this regime assumes that for free in-plane motion there are many available states to which an electron may scatter during interactions with helium atoms~\cite{Saitoh_1977,monarkha2013two}. This condition will likely change for a single electron with strongly confined in-plane motion leading to a confinement-induced suppression of the relaxation rate. In future experiments, the use of resonators made from high-kinetic inductance materials, which increase electron-resonator coupling, along with improvements in the measurement readout scheme, could enable the extraction of single-electron relaxation rates and facilitate investigation of electron scattering mechanisms at temperatures above 1 K.

\section*{Acknowledgements}
We thank J.~Theis for technical support. This work made use of the Pritzker Nanofabrication Facility of the Institute for Molecular Engineering at the University of Chicago, which receives support from Soft and Hybrid Nanotechnology Experimental (SHyNE) Resource (NSF ECCS-2025633), a node of the National Science Foundation’s National Nanotechnology Coordinated Infrastructure.

\appendix

\section{Appendix: Device and Experiment}
\label{sec:device}

The device is fabricated on a high-resistivity silicon wafer section using multi-layer electron-beam lithography and chlorine plasma etching. A set of electrodes (niobium, thickness 80 nm), including the resonator electrode, are patterned on to the substrate before the silicon is selectively etched to create a system of trenches some $d = 700$~nm deep. A second set of electrodes (niobium, 30~nm) are then patterned at the bottom of these trenches. In the final fabrication step aluminum air bridges~\cite{chen2014fabrication}, each several tens of $\mu$m in length and several $\mu$m in height, positioned far (at least 100~$\mu$m) from the trench network are used to make necessary electrode interconnects.

As shown schematically in Fig.~\ref{fig:device}(a), this process allows the fabrication of microchannel structures in which the electrostatic potential is controlled by the integrated electrodes. In our experimental cell, which contains a small reservoir containing 0.2 cc of liquid helium $\sim$2 mm beneath the device, a thin ($\sim$~30 nm) film of superfluid helium covers all internal surfaces including the upper surface of the device. This film allows helium to flow into the microchannels which then fill completely with the liquid due to capillary action. The liquid surface is then charged with electrons emitted from a small tungsten filament positioned several mm above the device. Each electron is attracted to the helium by its own image charge but is unable to enter due to a $\sim$1~eV potential barrier at the surface. As a result the electron `floats' some 10~nm above the helium surface separated from its neighbors by typically $d_{\mathrm{e}}\sim$100~nm due to mutual Coulomb repulsion. The surface electron density $n_s$, and so the spacing between electrons, can be controlled by the bias voltages applied to the device electrodes up to an electrohydrodynamic instability which typically occurs at densities close to $10^{10}$~cm$^{-2}$ in micron-scale channels. 

As is common in microfabricated surface-electron trapping devices, we employ an extended array of microchannels (28 channels, each 5~$\mu$m wide and 600~$\mu$m long) connected in parallel to create a reservoir region in which some $\sim$10$^6$ electrons are stored~\cite{glasson2001observation}. The presence of electrons in the reservoir, following emission from the filament, is confirmed by traditional Sommer-Tanner measurements using electrodes patterned on the base of the reservoir microchannels~\cite{sommer1971mobility, rees2016structural}. The reservoir is connected to the electron trap region by a 3~$\mu$m-wide channel, length 100~$\mu$m, at the bottom of which is the Bottom electrode. This channel is partially shown in Fig.~\ref{fig:device}(b). The Bottom electrode is typically biased positive to allow electrons to flow from the reservoir and collect above it. Thus, the reservoir microchannels and the microchannel leading to the trap are populated with a continuous sheet of electrons with electrochemical potential equal to, or close to, the ground plane potential (0~V). The electron density in the microchannel is given to good approximation by a parallel plate capacitor model as $n_s = \varepsilon \varepsilon_0 V_{\mathrm{b}} / ed$ where $\varepsilon = 1.057$ is the dielectric constant of liquid helium and $\varepsilon_0$ is the vacuum permittivity.

Microwave signals were transmitted to the cold stage of the cryostat through stainless steel coaxial cables and, after passing through a -20 dB cryoattenuator and a directional coupler (KRYTAR 102008020) providing a further 20 dB signal attenuation, entered the sample cell and were coupled to the resonator via the on-chip coupling port. The reflected microwave signal was amplified at room temperature with two low-noise amplifiers (Mini-Circuits ZX60-83LN-S+) giving a total +40 dB signal amplification. A vector network analyzer was used to perform the $S_{11}$ reflection measurements with a microwave output power $P_{\mathrm{VNA}} = -35$~dBm. Each dc bias line was filtered using on-chip LC filters with cut-off frequencies close to 1 GHz. 

The device was wire-bonded to a FR4 printed circuit board (PCB), which was then hermetically sealed with indium wire inside the copper cell. Electrical connection to the PCB was made via mini-SMP connectors. The cell was mounted on the cold stage of the cryostat. Once at base temperature, helium was admitted to the cell from room temperature via a 1.6 mm OD stainless steel tube thermally anchored at the 50 K, 4 K and 1 K stages of the cryostat. 

\section{Appendix: Resonator Model}
\label{sec:resonator}

As shown schematically in Fig.~\ref{fig:device}(b) the resonator is a $\lambda/2$ CPW resonator of total length $l=8.72$~mm consisting of two arms, which can be viewed as two quarter-wavelength resonators capacitively coupled at the trap location and galvanically connected at the other end (point \textit{c}). The width of the resonator electrode is 2~$\mu$m and the gap to the surrounding ground plane (not depicted in Fig.~\ref{fig:device}(b)) is 10~$\mu$m. The capacitive coupling at the open ends of the CPW lines hybridizes the two $\lambda/4$~-modes. This results in the formation of common (+) and differential (-) modes with ac voltage eigenvectors $\phi_{\pm} = (\phi_{a} \pm \phi_{b})/\sqrt{2}$~\cite{koolstra2024high}. Here $\phi_a$ and $\phi_b$ are ac voltages defined at the open ends (points \textit{a} and \textit{b} in Fig.~\ref{fig:device}b) of the CPW lines, as shown in the lumped-element schematic model of the coupled resonator-electron system depicted in Fig.~\ref{fig:s11}(a). For the differential mode, an ac voltage across the trap oscillates out-of-phase, generating an electric field $\mathbf{E}$ that excites the electron motion, thereby coupling to the electron system. The frequency of the differential mode is given by $f_r = \omega_{r}/2\pi = 1/2\pi \sqrt{L_rC}$, where $C = C_r + 2C_{\mathrm{dot}}$ is the effective capacitance of the mode and the values $C_r = 0.24$~pF and $L_r = 2.8$~nH were determined using conformal mapping techniques with a small ($\times 1.3$) adjustment applied to $L_r$ to better fit the measured frequencies. The cross-capacitance between the two arms of the resonator center pin electrodes $C_{\mathrm{dot}}$ is primarily governed by the geometry of the trap. From FEM calculations, the extracted value of $C_{\mathrm{dot}} = 20$~aF ($\ll C_r$) has a negligible impact on the resonance frequency. The common mode frequency $f_c=\omega_{c}/2\pi = 1/2\pi \sqrt{1/(L_r + 2L_t)C_r} < 4$~GHz is suppressed by engineering a large tail inductance $L_t$ using a long CPW line of length $\sim 2.3$~mm connected at point \textit{c}~\cite{koolstra2024high}. To bias the resonator, the voltage $V_{\mathrm{r}}$ is applied at the opposite end of the tail CPW line. 

The resonator is measured in reflection, with capacitive coupling to the drive port through a small section of one of the resonator arms, as illustrated in Fig.~\ref{fig:device}(b). The reflection coefficient $S_{11}$ near the resonance can be expressed as:
\begin{equation}
    S_{11} = 1 - \frac{2 Q_t}{Q_c} \frac{a e^{i \theta}}{1 + i 2 Q_t \delta},
    \label{eq:reflection}
\end{equation}
where $Q_t = ( 1/Q_i + 1/Q_c)^{-1}$ is the total quality factor, $Q_i$ and $Q_c$ are the internal and coupling quality factors respectively, $\delta = (f - f_r)/f_r$, and the complex quantity $a e^{i \theta}$ accounts for asymmetries of the resonator's reflection due to impedance mismatches in the measurement circuit. A typical measured reflection spectrum displays a dip in magnitude near the resonance frequency as shown in Fig.~\ref{fig:loading}(b). Fitting to this data gives the resonance frequency of the differential mode  $f_r = 6.04383$~GHz with internal quality factor $Q_i = 5800$ and coupling quality factor $Q_c = 4800$. These measurements were performed at $T=1.1$~K before charging the helium surface. We note that the experiments shown here were successfully repeated with several devices, each with different but quantitatively similar values of $f_r$ and $Q_i$.

\section{Appendix: Electron-Resonator Coupling Model}
\label{sec:ehe-res_coupling}

\begin{figure}
\includegraphics[scale=0.24]{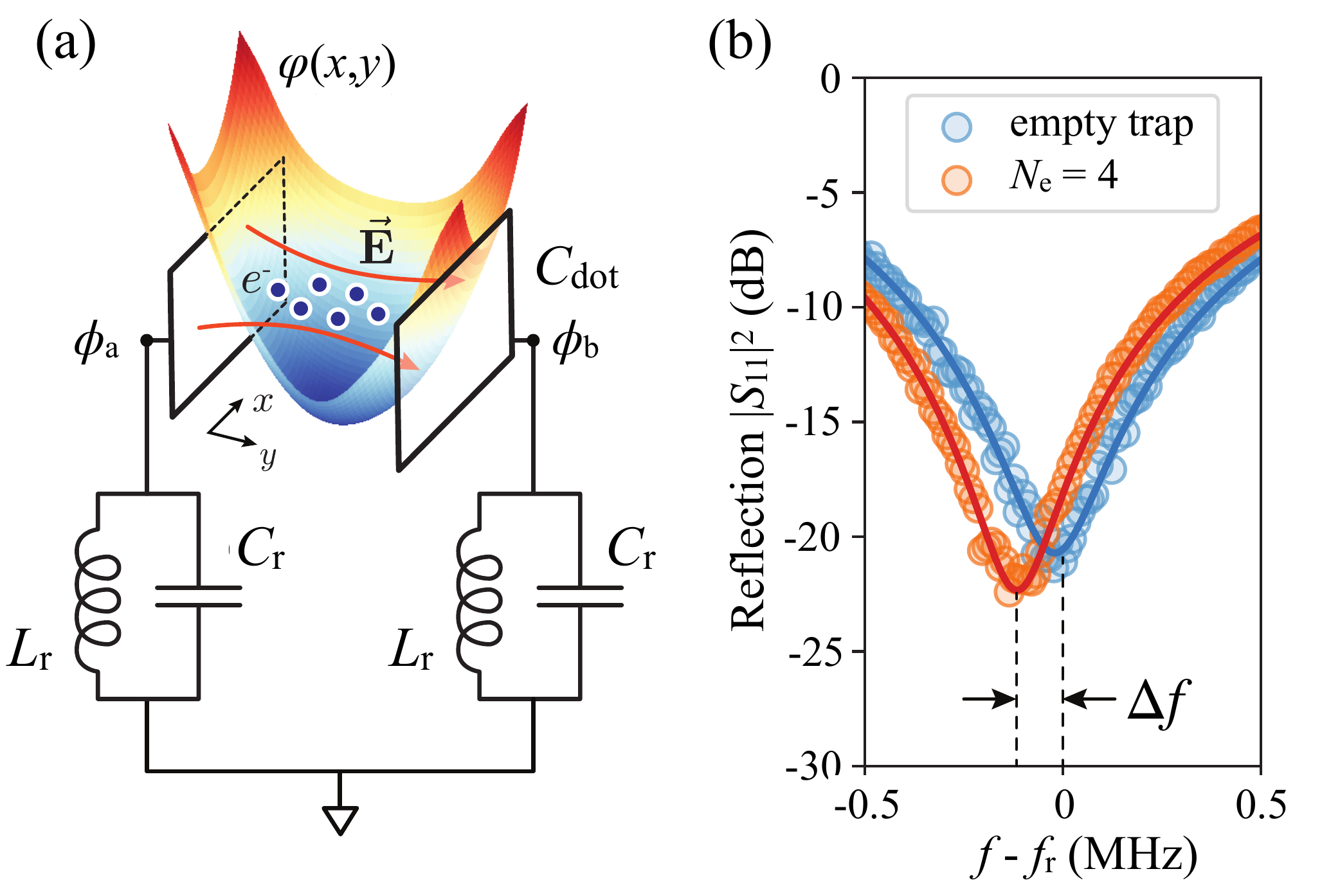}
\caption{\label{fig:s11} (a) Schematic diagram of the circuit model used to describe the electron-resonator coupling as discussed in the main text. FEM is used to calculate the potential well which traps the electrons between the resonator arms, which are modeled as capacitor plates. (b) Experimental data showing the frequency shift $\Delta f$ measured when loading the trap with $N_e = 4$ electrons. The solid lines represent fitting curves obtained using Eq.~\ref{eq:reflection}.}
\end{figure}

In our experiments, electrons are loaded into an electrostatic trap created at the open ends of the resonator electrodes. The trap potential $\varphi(x,y)$ is engineered to facilitate motion along the direction of the resonator electric field, in this case the $y$-axis. The trap potential $\varphi(x,y) = \sum_{k} \alpha_k(x,y) V_k$ is determined by dimensionless functions $\alpha_k$ representing the potential due to electrode $k$ at unit voltage, which we calculate from FEM. For the temperatures relevant to our experiments, where $k_{\mathrm{B}} T > \hbar \omega_e$ (with $\omega_\mathrm{e}$ representing the characteristic single electron motion frequency determined by trap potential curvature), we describe the non-degenerate electron system using classical equations of motion in a long-wave approximation. We assume that the
anisotropy of the trapping potential allows the system to be treated as one-dimensional along the $y$-axis. Electron motion is excited by the electric field generated by the ac voltage $(\phi_{a} \pm \phi_{b})/\sqrt{2}$ on the capacitor plates (see Fig.~\ref{fig:loading}a). We can express the linearized equations of motion (EOM) for the $i^{\mathrm{th}}$ electron displacement $y_i$ around its equilibrium position as follows:
\begin{equation}
    \Ddot{y}_i + \omega_i^2 y_i + \sum_{j \neq i} \kappa_{ij}y_j = -e E_i e^{i \omega t}/m_e.
    \label{eq:eom}
\end{equation}
Here, $\omega$ represents the frequency of the microwave probe pump, $\omega_i$ denotes the frequency of motion for the $i^{\mathrm{th}}$ electron near its equilibrium position 
, $e$ is the elementary charge, $m_e$ is the electron mass, and $E_i$ is the electric field generated by the resonator electrodes at the electron location. Coulomb interactions between electrons are approximated by the linear term $\kappa_{ij} = e^2/2 \pi \epsilon_0 m_e d_{ij}^3$, where $d_{ij}$ represents the equilibrium distance between electrons. We note that the coupling term $\kappa_{ij}$ represents unscreened Coulomb interactions, which can be reduced in our trap geometry due to screening effects from nearby electrodes. The response of the electron system to perturbing microwave fields is characterized by its linear electric susceptibility (or ac current susceptibility) $\chi_e(\omega)$, which quantifies the system's polarization. This quantity is found by solving the system of Eq.~\ref{eq:eom} (see derivation in Appendix~\ref{sec:a_solvingeom}) expressed as: 
\begin{equation}
    \label{eq:susceptibility} 
    \chi_e(\omega) = \sum_{n} \frac{4g_n^2}{\omega_n^2 - \omega^2 + 2i\omega\Gamma_n},    
\end{equation}
where $g_n$ represents the coupling strength between the microwave field and the $n^{\mathrm{th}}$ \textit{vibrational} mode of the electron system, and $\omega_n$ is the eigenfrequency of that mode. $\Gamma_n$ is the phenomenological damping constant which describes the energy relaxation due to interactions between the electron and $^{4}$He vapor atoms, quantized He surface capillary waves (ripplons), and bulk phonons~\cite{monarkha2013two}. The coupling strength $g_n$ determines the dipole energy of the vibrational mode and is defined by the dot product of the field strength vector $\vec{\mathbf{\mathcal{E}}} = \grad{\vec{\alpha}}$ at the electron equilibrium locations (defined as per unit voltage) and the normalized eigenvector $\vec{\mathbf{x}}_n$ of the specific vibrational mode:
\begin{equation}
    g_n = \frac{e(\vec{\mathbf{\mathcal{E}}} \cdot \vec{\mathbf{x}}_n)}{2\sqrt{m_eC}}.
    \label{eq:gn}
\end{equation}
With these definitions of $g_n$ and $\vec{\mathbf{\mathcal{E}}}$, Eq.~(\ref{eq:susceptibility}) generalizes to the 2D and even 3D bound case. We note that in a uniform electric field and without interactions, the coupling strength is given by $g_n = \sqrt{N} g_\mathrm{e}$, where $g_\mathrm{e}$ represents the coupling energy for a single electron, which exhibits the $\sqrt{N}$ scaling characteristic of the Dicke and Tavis-Cummings model~\cite{taviscummings,blaha2022beyond}.

\begin{figure}
\includegraphics[scale=0.75]{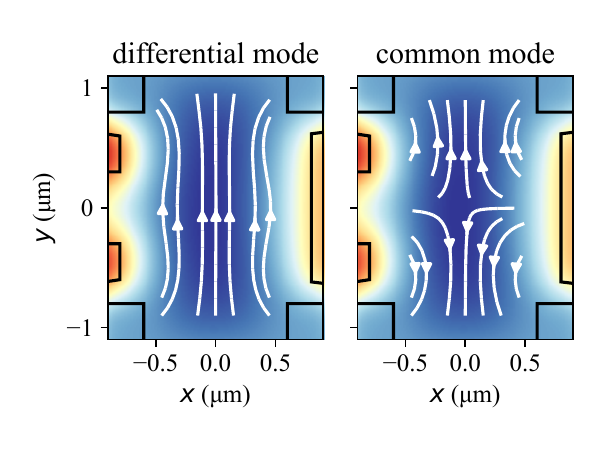}
\caption{\label{fig:fieldprofile} Electric field lines (white arrows) in the trap region created by the differential (left) and common (right) modes of the resonator.}
\end{figure}

The expression for the coupling also indicates which vibrational modes can be excited by a particular mode of the resonator, as the field vector depends on the specific resonator mode being driven. This relationship can be readily deduced from the symmetries of the vibrational mode of the electron system and the resonator modes. Electric field distributions of the resonator modes are shown in Fig.~\ref{fig:fieldprofile}. For the differential and common modes of coupled LC circuits, each element in the vector $\vec{\alpha}$ is determined by the dot product of the resonator electrode coupling constants $\icol{\alpha_a\\\alpha_b}$ evaluated at the electron positions with the eigenvectors of the resonator modes $\icol{1\\\pm 1}/\sqrt{2}$. Here $\alpha_{a,b}$ are the coupling constants for the resonator's upper and lower arms, respectively. Equation~\ref{eq:susceptibility} clearly shows that the largest response in the frequency shift will be induced by vibrational modes where the eigenfrequencies are closest to the resonator frequency and the dipole coupling $g_n$ is large.

The response of a single electron in a symmetrical trap can be readily derived using Eq.~(\ref{eq:susceptibility}) and~(\ref{eq:gn}). Single electron motion in the trap is solely determined by the electrostatic trap curvature $\omega_e^2 = e m_e^{-1} \big [ d^2\varphi(x,y) / dy^2 \big ]_{x,y=0}$, which is controlled by gate voltages. The coupling strength simplifies to $g_\mathrm{e} = \sqrt{2} e \mathcal{E}_y / 2 \sqrt{m_e C}$, which coincides with the expression for the dipole energy $e \mathcal{E}_y\hat{x}\hat{V}$ of zero-point fluctuations of the electron motion $\hat{x}$ and vacuum fluctuations of the resonator field $\mathcal{E}_y\hat{V}$. Here $\mathcal{E}_y = (\nabla_y \alpha_a - \nabla_y \alpha_b)/2$ is the electric field strength per unit voltage created by the differential mode of the resonator at $y=0$. We note that driving electron motion with the differential mode gives $\sqrt{2}$ enhancement in the coupling compared to a single-mode resonator. For the trap design shown in Fig.~\ref{fig:device}c we find $\mathcal{E}_y = 0.23$~$\mu$m$^{-1}$ which gives single electron-cavity photon coupling energy $g_\mathrm{e}/2\pi = 9.8$~MHz. For these calculations, we used Eq.~\ref{eq:gn} with a slight adjustment to the resonator capacitance, $\gamma C$, where $\gamma$ serves as a fitting parameter. We find that $\gamma = 0.85$ provides good agreement with the experimental data for the single and multi-electron frequency shifts presented in Fig.~\ref{fig:multielectron} and Fig.~\ref{fig:singleelectron}. It is evident that for the common mode $\mathcal{E}_y = (\nabla_y \alpha_a + \nabla_y \alpha_b)/2 = 0$ at the center of the trap due to it's symmetry (see Fig.~\ref{fig:fieldprofile}). Consequently, this resonator mode cannot excite single electron motion. For larger numbers of electrons in the arbitrary trap potential $\varphi(x,y)$ we determine the eigenmodes and equilibrium configuration of interacting electrons by numerical minimization procedures~\cite{koolstra2024high,quantumelectron}.
Figure~\ref{fig:s11}b shows measured spectra both without electrons and with $N_e = 4$ electrons in the trap. The observed frequency shift, $\Delta f \approx -100$~kHz, agrees well with the numerically calculated value of $\Delta \omega / 2\pi = -95$~kHz.

\section{Appendix: Solving EOM}
\label{sec:a_solvingeom}

We look for solutions of Eq.~\ref{eq:eom} that are periodic in time with angular frequency $\omega$, i.e. $\vec{y}(t) = \vec{y}_0 \exp(i\omega t)$, where $\vec{y}_0 = (y_1^0, y_2^0, \ldots, y_N^0)^T$. Substituting this ansatz into Eq.~\ref{eq:eom}, the coupled equations of motion can be expressed in the matrix form as $(\mathbb{M} - \omega^2 \mathbb{I}) \vec{y}_0 = -e\vec{E}/m_e$, where damping terms are omitted for simplicity, and the matrix $\mathbb{M}$ is defined as:
\begin{equation}
    \mathbb{M} =
    \begin{pmatrix}
      \omega_1^2 & \kappa_{12} & \cdots & \kappa_{1N} \\
      \kappa_{21} & \omega_2^2 & \cdots & \kappa_{2N} \\
      \vdots  & \vdots  & \ddots & \vdots  \\
      \kappa_{N1} & \kappa_{N2} & \cdots & \omega_N^2 
    \end{pmatrix}
\end{equation}
Representing the solution in terms of the eigenvector basis $\vec{x}_n$ of the unperturbed electron system as $\vec{y}_0 = \sum_{n} c_n \vec{x}_n$, with eigenfrequencies defined as $\mathbb{M} \vec{x}_n = \omega_n^2 \vec{x}_n$ we obtain the solution in the form
\begin{equation}
    \vec{y}_0 = -\frac{e}{m_e} \sum_n\frac{E_n}{\omega_n^2 - \omega^2}\vec{x}_n, 
\end{equation}
where $E_n = (\vec{E}\cdot \vec{x}_n)$ represents an effective electric field driving the $n^{\mathrm{th}}$-vibrational mode of the electron system. Next, we define the polarization density projected onto the electric field vector as $P = -e \mathcal{V}_{m}^{-1} (\vec{y}_0\cdot\vec{E}/E_0)$, where $\mathcal{V}_m$ is the electric field mode volume of the specific resonator mode, and $E_0$ represents the characteristic field strength associated with the drive voltage amplitude $V_0$. By expressing the mode volume in terms of the resonator capacitance as $\mathcal{V}_m = CV_0^2/\epsilon_0 E_0^2$ we can rewrite polarization density in a linear response form $P = \epsilon_0 \chi_e(\omega) E_0$, where the electric susceptibility $\chi_e(\omega)$ is given by Eq.~\ref{eq:susceptibility_main} in the main text.

\section{Two Electron Motional Modes.}
\label{sec:a-twoelectron}
Here we present the derivation of analytical expressions for a system of two electrons coupled to a resonator field. We start with the Hamiltonian describing a 1D motion of two electrons within a trapping potential:
\begin{equation}
    \mathcal{H}_e = \sum_{i=1,2} \Big [ \frac{m \dot{x_i}^2}{2} - e\varphi(x_i) \Big ] + \frac{e^2}{4 \pi \epsilon_0} \frac{1}{|x_1 - x_2|}
    \label{eq:2electron_hamiltonian}
\end{equation}
The equilibrium positions of the electrons $x_0$ are determined from the force balance equation $\partial \varphi / \partial x \big|_{x=x_0} = e / (4 \pi \epsilon_0 d^2)$, where $d = 2x_0$ is the distance between the electrons. Subsequently, we perform a Taylor expansion of both the potential energy and the Coulomb energy in terms of the displacements \(\delta x_i\) around the equilibrium positions, which results in the Hamiltonian:
\begin{equation}
    \begin{aligned}
        &\mathcal{H}_e = \sum_{i} \Big( \frac{m \dot{\delta x_i}^2}{2} + \frac{m\omega_e^2}{2} \delta x_i^2 \Big) + \frac{m\omega_{C}^2}{2} \Big( \delta x_1 - \delta x_2 \Big)^2,
        \label{eq:2electron_hamiltonian_disp}
    \end{aligned}
\end{equation}
where $\omega_C^2 = 2e^2/4 \pi \epsilon_0 m d^3$ and we have used the relation between the curvature of the potential and the single-electron frequency $\partial^2 \varphi / \partial x^2 \big|_{x = \pm x_0} = -m \omega_e^2 / e$ valid in the harmonic approximation of the trap potential. We have also omitted constant terms, and the linear terms in displacement cancel out due to the force balance condition. This Hamiltonian can be diagonalized using the following transformation:
\begin{equation}
    \left\{
    \begin{aligned}
        &x_{+} = (\delta x_1 + \delta x_2)/\sqrt{2} \\
        &x_{-} = (\delta x_1 - \delta x_2)/\sqrt{2}
        \label{eq:electron_commdiffmodes}
    \end{aligned} \right.
\end{equation}
The new coordinates correspond to two vibrational modes, which are the in-phase (+) and out-of-phase (-) motion with eigenvectors $\vec{\mathbf{x}}_{\pm} = \icol{1\\\pm 1}/\sqrt{2}$. As $x_+^2 + x_-^2 = \delta x_1^2 + \delta x_1^2$, the two electron Hamiltonian takes the final form:
\begin{equation}
    \mathcal{H} = \frac{m \dot{x}_{+}^2}{2} + \frac{m \omega_{2+}^2 x_{+}^2}{2} + \frac{m \dot{x}_{-}^2}{2} + \frac{m \omega_{2-}^2 x_{-}^2}{2},
    \label{eq:2electron_h_norm}
\end{equation}
where the frequency of the in-phase and out-of-phase motion is given by $\omega_{2+}^2 = \omega_e^2$ and $\omega_{2-}^2 = \omega_e^2 + 2\omega_{\mathrm{C}}^2$, respectively.

 The in-phase electron motion couples to the differential mode, which is evident from the nonzero value of the dot product $(\vec{\mathbf{\mathcal{E}}} \cdot \vec{\mathbf{x}}_{+})$ with electric field strength vector defined as $\vec{\mathbf{\mathcal{E}}} = \sqrt{2} \mathcal{E}_{y,2} \icol{1\\1}$. 
In this case both electrons experience an equal field strength $\mathcal{E}_{y,2} = \nabla_y \big [ (\alpha_a - \alpha_b)/2 \big ]_{y = d/2}$ aligned with their displacement vectors. In the harmonic approximation this results in a two electron-resonator coupling energy $g_{2+} = \sqrt{2} g_\mathrm{e}$. This leads to values of $\Delta f$ that are twice larger than the single electron case. From the dot product $(\vec{\mathbf{\mathcal{E}}} \cdot \vec{\mathbf{x}}_{-})$ it is clear that the out-of-phase motion of two electrons can only be excited by the common mode of the resonator.

\bibliography{SingleElectron}

\end{document}